\documentclass[
  ,final            
  ]
  {aipproc}

\layoutstyle{6x9}

\newcommand{\beqn}{\begin{equation}}
\newcommand{\eeqn}{\end{equation}}
\newcommand{\barr}[1]{\begin{array}{#1}}
\newcommand{\earr}{\end{array}}
\newcommand{\beqna}{\begin{eqnarray}}
\newcommand{\eeqna}{\end{eqnarray}}

\newcommand{\gapproxeq}{\lower.7ex\hbox{$\;\stackrel{\textstyle>}
{\sim}\;$}}
\newcommand{\lapproxeq}{\lower.7ex\hbox{$\;\stackrel{\textstyle<}
{\sim}\;$}}
\newcommand{\plabel}[1]{\label{#1}}
\newcommand{\pbibitem}[1]{\bibitem{#1}}

\begin{document}

\title{
The isotensor pentaquark\footnote{Expanded version of a talk presented
at the X International Conference on Hadron Spectroscopy (HADRON '03), 
31 Aug. - 6 Sept., Aschaffenburg, Germany.}}

\author{Philip R. Page}{
  address={Theoretical Division, MS B283, Los Alamos
National Laboratory, Los Alamos, NM 87545, USA}
}

\begin{abstract}
Further consequences of the 1540 MeV $\Theta^+$ resonance as an
isotensor pentaquark beyond Capstick {\it et al.}~\cite{cpr} are
explored. It is argued that the SAPHIR data may not currently 
exclude the existence of the charged partner $\Theta^{++}$.
The usual prediction of the dominance of non-resonant $\Theta^+ K$, and 
$\Theta^+ K^\ast$,
final states in photoproduction on the proton is argued not to obtain
for an isotensor $\Theta^+$. This enhances the importance of excited
baryon final states, where the excited baryon decays to $\Theta^+ K$ or
$\Theta^+ K^\ast$;
as well as the non-resonant $\Theta^+ K \pi$ final state. 
The small width of the recently discovered $\Xi^{--}$ cascade resonance to
$\Xi^-\pi^-$ is easier to explain if $\Theta^+$ is an isotensor
pentaquark than if it is in the $\bar{\mbox{\bf 10}}$ 
representation, due to both
an isospin and U-spin selection rule. 
A new production diagram for $\Theta^+$ in the photoproduction on the
deuteron is suggested. 
\end{abstract}

\maketitle



\section{An isotensor pentaquark explains the $\Theta^+$ width}

The consensus of various experiments is that the total width $\Gamma$
of $\Theta^+$ is less than 9 MeV~\cite{cpr}. More restrictive bounds on the
width emerge from its non-observation in $K^+ d$ scattering ($\Gamma <
6$ MeV) \cite{nussinov} and $K^+$-nucleon scattering ($\Gamma \lapproxeq
1$ MeV)~\cite{arndt}. It was proposed that the narrowness of the
$\Theta^+$ can be explained if it is an isotensor state, in which case
the decay to the kinematically allowed channels $nK^+$ and $p K^0$ is
isospin violating~\cite{cpr}. Based on this hypothesis, an
upper bound of roughly $0.45$ MeV was put on the
width~\cite{cpr}, consistent with all experimental data above.
If $\Theta^+$ is isotensor, other charge states like $\Theta^{++}$ should
exist.

\section{SAPHIR may not have excluded the existence of $\Theta^{++}$}

In addition to the observation by SAPHIR of the $\Theta^+$ with $63
\pm 13$ events, they also see a statistically insignificant
$\Theta^{++}$ signal with $75\pm 35$ events in the reaction $\gamma p
\to \Theta^{++} K^-\to p K^+ K^-$~\cite{saphir}.  The SAPHIR detector 
appears to
have an acceptance\footnote{SAPHIR estimates an acceptance ratio of
$5000$ events $/$ $63$ events $/$ $(3$ to $4) \;\times$ $1/2$ $\times$
$2/3$ $=7.6$~\protect\cite{saphir}.}  that is about eight times higher
in $K^+ K^- p$ than in $K^+ K^0_S\, n$. An estimate for the ratio of
cross-sections for $\Theta^{++}$ and $\Theta^{+}$ production is
then\footnote{Noting that the observed $K^0_S$ is only produced half
the time from $\bar{K}^0$ and that the detected $\pi^+\pi^-$ mode of
$K^0_S$ has a branching ratio of about $2/3$.  The branching ratio
$Br(\Theta^{++}\to p K^+)$ is very close to unity if $\Theta^{++}$ is
below the $NK\pi$ threshold~\protect\cite{cpr}.}

\beqn\plabel{sap}
\frac{\sigma(\gamma p \to \Theta^{++}K^-)}
{\sigma(\gamma p \to\Theta^{+} \bar{K}^0)} =
\frac{1}{2}\;\frac{2}{3}\;\frac{1}{8} \frac{N(\Theta^{++})}{N(\Theta^{+}) } 
\frac{Br(\Theta^+\to n K^+)}{Br(\Theta^{++}\to p K^+)}
\leq \frac{1}{24}\;\frac{75\pm 35 \mbox{}}{63 \pm 13 \mbox{}} = 
\frac{1}{20 \pm 10} \; .
\eeqn
Thus the process $\gamma p \to \Theta^{++}K^-$ is at least 
$20\pm 10$ times weaker than $\gamma p \to\Theta^{+} \bar{K}^0$. 
The cross-section 
$\sigma(\gamma p \to\Theta^{+} \bar{K}^0)$ was measured to be 
200 nb~\cite{saphir}. A preliminary analysis by CLAS in the {\it same}
reaction with the {\it same} photon energy range does not see
the $\Theta^+$ and gives 
a cross-section $< 20$ nb~\cite{hicks}. If the CLAS result is correct, 
then the serious discrepancy
suggests that the SAPHIR analysis may well be in error, and the existence
of the $\Theta^{++}$ is not excluded.

The photoproduction of an isotensor $\Theta$ through the process
$\gamma p \to K \Theta$ cannot proceed via isospin conserving
interactions through the isoscalar component of the photon, so that
the process is taken to proceed through the isovector component
(usually associated with the $\rho^0$ via vector meson
dominance)~\cite{cpr} or the smaller isotensor component $T^0$ arising from
four-quark Fock states. If $\gamma
= c_0 |I=0\rangle + 
c_1 |I=1\rangle + c_2 |I=2\rangle$ the scattering T-matrix element

\beqna\plabel{iso}
\lefteqn{\hspace{-12cm}\langle\gamma p | T | \Theta K \rangle =
\sum_n \langle\gamma p | n \rangle \; \langle n | T | \Theta K \rangle = 
} & & \nonumber \\
c_1 \left\langle \frac{3}{2} \frac{1}{2}\: |\: 2 I^z_\Theta \:
\frac{1}{2} I^z_K 
\right\rangle \sum_n \langle \rho^0 p | n \rangle \; 
\langle n |T | \Theta K \rangle_R
+ c_2  \sum_n \langle T^0 p | n \rangle \; 
\langle n |T | \Theta K \rangle
\eeqna 
where a formal sum over asymptotic states $n$ has been inserted.
The Clebsch-Gordon coefficient has been isolated explicitly from the
$\langle n |T | \Theta K \rangle$ overlap corresponding to the 
$I=1$ photon component, noting
that only intermediate states with $I={3}/{2}$ can contribute
in this case, and assuming that strong interactions conserve
isospin. If the isotensor component of the photon is negligible,
as is usually assumed,
the amplitude ratio for scattering to $\Theta^+$ ($I^z_\Theta = 0$)
and $\Theta^{++}$ ($I^z_\Theta = 1$) is
$\langle \, \frac{3}{2}\frac{1}{2} \, | 20\, \frac{1}{2}\frac{1}{2}\, 
\rangle\; / \;\langle \, \frac{3}{2}\frac{1}{2} | 21\, 
\frac{1}{2}-\frac{1}{2}\, \rangle = -\sqrt{\frac{2}{3}}$. The
$\Theta^{+}$ is produced with a cross-section $2/3$ that of
$\Theta^{++}$ (the SAPHIR calculation obtained a factor of
$1/3$~\cite{saphir}).
In Eq.~\ref{sap} this ratio was estimated to be $20\pm 10$,
which led SAPHIR to conclude that $\Theta^{++}$ does not exist~\cite{saphir}.

\section{Isotensor $\Theta^+$ photoproduction on the nucleon}

Photoproduction of $\Theta^+$ on the nucleon is typically calculated in
hadronic models because perturbative QCD is not
applicable~\cite{photon}. However, in any of these models diagrams are
missing, e.g. the {\it a priori} important proton sea 
diagram in Fig.~\ref{prodfig},
and there is a proliferation of unknown coupling constants.
It is therefore instructive to return to a na\"{\i}ve quark level
discussion understanding that the diagrams are not those of
perturbative QCD, but represent Green's functions needed to evaluate
the scattering T-matrix. The discussion will assume that there is
a penalty for each time a quark-antiquark pair needs to be created from 
the vacuum.

Consider the generic process $\gamma N \to K^\star \Theta^+$ for an
isotensor $\Theta^+$. Here $K^\star$ represents either the ground
state $K$ or an excited state with the flavor of $K$. In order for the
process to happen, one light quark $n\bar{n}\equiv
(u\bar{u}+d\bar{d})/\sqrt{2}$ pair and one $s\bar{s}$ pair must be
created.  The dominant production process is expected to be where one
of the pairs that need to be created is created by the photon.  (Such
production processes can be found in the T-channel mechanisms
suggested in Fig. 4 of Ref.~\cite{stepanyan}, Fig. 1 of
Ref.~\cite{saphir} and Refs.~\cite{photon}.)  Because the isoscalar
component of the photon does not contribute within isospin conserving
interactions, this can only be the $n\bar{n}$ pair.  The $s\bar{s}$
pair will then be created from gluons. The three light quarks in the
incoming $N$ will directly go into the outgoing $\Theta^+$. However,
this is isospin forbidden: the three light quarks in the $\Theta^+$ is
isospin $3/2$, and the $N$ is isospin $1/2$.
Since the would-be dominant production process vanishes within isospin
symmetry, the production of the $\Theta^+$ is more complicated. The
photon interacts with a quark or antiquark, but the two $q\bar{q}$
pairs are both created from gluons. 

\begin{figure}
\leavevmode
\includegraphics[height=.18\textheight]{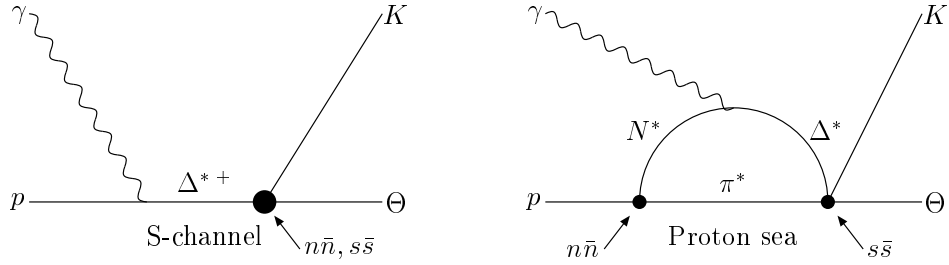}
\caption{\plabel{prodfig} S-channel and proton sea production.
An asterisk indicates an off-shell particle with the flavor
structure of the corresponding on-shell particle. The site of pair creation
is indicated by a filled circle, and the pair(s) that is created is shown.
In the S-channel diagram the photon interacts first, 
and the two pairs are created later. In the proton sea diagram 
the light quark $n\bar{n}$ pair is created first, the 
photon subsequently interacts, and the strange quark 
$s\bar{s}$ pair is created last.
}
\end{figure}

$\gamma N \to K \Theta^+$~\cite{saphir}: 
For this process intermediate resonance
mechanisms of the type $\gamma N \to \Delta^\star \to K \Theta^+$
should be considered (similar to the S-channel diagram in 
Fig.~\ref{prodfig}). 
(An intermediate $N^\star$ is not allowed by
isospin conservation.) The production process $\gamma N \to
\Delta^\star$ requires no pair creation, while the decay $\Delta^\star
\to K \Theta^+$ requires two pair creations, and has to compete with
the various decay modes to a baryon and meson.  There are also
non-resonant mechanisms which require two pair creations, for example
the proton sea diagram in Fig.~\ref{prodfig}.  Hence the
$\gamma N \to \Delta^\star$ and non-resonant $K \Theta^+$ mechanisms
are competitive.

$\gamma N \to K \Theta^+\pi $~\cite{stepanyan}:
Since the $\gamma N \to K^\ast
\Theta^+$ process requires two pair creations from the isospin
selection rule explicated above, other processes become
competitive. (The branching ratio $K^\ast \to K \pi$ is close to unity
so that the pair creation needed for the decay incurs no
penalty). Generally the process $\gamma N \to K \Theta^+ \pi$ can
happen by creating two $n\bar{n}$ pairs and one $s\bar{s}$ pair. This
can only be competitive with the $\gamma N \to K^\ast \Theta^+$
process if the photon creates one of the pairs.  Processes with
intermediate resonances of the type $\gamma N \to \Delta^\star \pi$
with $\Delta^\star\to K \Theta^+$ can happen via the photon creating
an $n\bar{n}$ pair, with $n$ going into the $\Delta^\star$, and
$\bar{n}$ going into the $\pi$. Such a process involves two pair
creations from the vacuum in the decay $\Delta^\star\to K \Theta^+$.
Another possibility is $\gamma N \to \Delta^\star$ where $\Delta^\star
\to K^\ast \Theta^+$ which again requires two pair creations for the
decay. There are also non-resonant mechanisms which require two pair
creations from the vacuum. (There are even non-resonant mechanisms
coming from a four-quark Fock component in the photon, including a possible
isotensor component, with {\it one} pair creation from the vacuum.)  Hence
the $\gamma N \to \Delta^\star \pi$, $\Delta^\star$, and non-resonant
$K^\ast \Theta^+$ and $K \Theta^+\pi $ mechanisms are competitive.

\section{The $\Xi^{--}$ and isotensor $\Theta^+$ can be consistent}

The recently discovered $\Xi^{--}$~\cite{na49} can be put in 
the same $SU_F(3)$ multiplet as the $\Theta^+$ if both are pentaquarks. 
An isotensor
$\Theta^+$ can only be put in a {$\bf 35$} representation of $SU_F(3)$
(mentioned in Ref.~\cite{harari}),
while an isoscalar $\Theta^+$ can only be put in the
$\bar{\mbox{\bf 10}}$ representation. Both these representations
also admit $\Xi^{--}$. In both representations
the $\Theta^+$ and $\Xi^{--}$ are in the same V-spin multiplet. V-spin
is an exact quantum number if $SU_F(3)$ is an exact symmetry of QCD.
The $p$ and $\Xi^{-}$ are in the same V-spin multiplet and
$SU_F(3)$ representation. Similarly for the $K^0$ and $\pi^-$.
Hence the decay amplitude $\Theta^+\to p K^0$ and that of
$\Xi^{--}\to \Xi^{-} \pi^-$ are related by a V-spin Clebsch-Gordon 
relation. If the $\Theta^+$ is isotensor the decay amplitude to $p K^0$ is
zero within isospin symmetry. The V-spin relation implies that the
decay amplitude $\Xi^{--}\to \Xi^{-} \pi^-$ is zero within $SU_F(3)$ symmetry.
If the $\Theta^+$ is isoscalar neither the $\Theta^+$ nor the
$\Xi^{--}$ decay amplitudes are zero by these symmetry arguments.

The decay $\Xi^{--} \to \Xi^{-} \pi^-$ is also suppressed by a U-spin
selection rule if $\Theta^+$ is isotensor, but not if it is isoscalar.
U-spin is an exact quantum number if $SU_F(3)$ is an exact symmetry of QCD.
The U-spin of $\Xi^{--}$ is $2$ and $0$ in the {$\bf 35$} 
and $\bar{\mbox{\bf 10}}$ representations respectively. The 
U-spin of $\Xi^{-}$ and $\pi^-$ is $1/2$. The decay 
$\Xi^{--} \to \Xi^{-} \pi^-$ is hence U-spin forbidden only if 
$\Xi^{--}$ is in the {$\bf 35$} representation, noting that this
``fall-apart'' decay proceeds without quark pair creation, i.e.
the interaction is a U-spin singlet. 

The fact that the decay $\Xi^{--}\to \Xi^{-} \pi^-$ is suppressed by
two independent symmetry arguments if $\Theta^+$ is isotensor goes a
long way towards explaining the small $<18$ MeV total width of 
$\Xi^{--}$~\cite{na49}. The fly in the ointment is that $\Xi^{--}$
can also decay to $ \Xi^{\ast\, -} \pi^-$ and $\Sigma^- K^-$ via 
fall-apart decay, and to $\Xi\pi\pi$ and $\Sigma K\pi$ via one vacuum 
$n\bar{n}$ pair creation.
It remains to be explained why these decay widths are small.

\section{Photoproduction through the $\pi$ in the deuteron}

In the reaction $\gamma d \to p (n) K^+ K^-$ studied at CLAS~\cite{clas}
the $p$ and $K^-$ must be detected in order to reconstruct the final
state. If the $p$ is a spectator, it will not be seen due to its
small kinetic energy~\cite{clas}. Only $K^-$ which are not produced forward
can be detected. Hence diagrams were suggested that are not of a spectator
nature~\cite{clas,schumacher}. The $K N \to \Theta^+$ fusion diagram
of Ref.~\cite{schumacher} is not allowed for an isotensor $\Theta^+$ due to
isospin conservation. The diagram originally suggested 
involved a $\gamma n \to \Theta^+ K^-$ vertex~\cite{clas}, which was shown
above to require two vacuum pair creations for an isotensor $\Theta^+$, 
and a $K^-$ rescattering on the photon. In Fig.~\ref{deut} a diagram is
displayed which requires two vacuum pair creations and no
rescattering, and should hence be dominant. The diagram is natural because
the deuteron is an extended nucleus mainly bound by 
long-distance $\pi$-exchange.

\begin{figure}
\leavevmode
\includegraphics[height=.13\textheight]{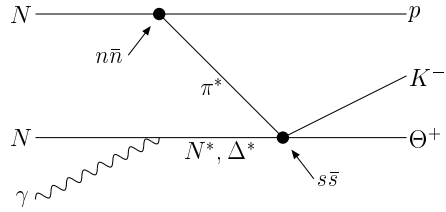}
\caption{\plabel{deut} Photoproduction of $\Theta^+$ on the deuteron. 
The neutron and proton in the deuteron 
are denoted by $N$. There are diagrams for
each possible assignment of the neutron and proton to the label $N$.
The other conventions are as is Fig.~\protect\ref{prodfig}.
}
\end{figure}

\section{Acknowledgments}

The help of both W.~Roberts and S.~Capstick (construction of the 
{\bf 27} and {\bf 35}
multiplets), W.~Roberts (noting the isotensor component of the photon),
D.~Weygand (noting the excited baryon production), and detailed
discussions with L.~Guo, K.~Hicks  and D.~Weygand on their experimental
data, are acknowledged. Additional helpful discussions with B.~L.~Berman,
P.~Bicudo, B.~Cahn, A.~Dolgolenko, T.~Goldman, E.~Klempt, H.~J.~Lipkin,
C.~A.~Meyer, R.~Schumacher, S.~Sasaki, S.~Stepanyan and M.~Wagner are
gratefully acknowledged.  This research is supported by the
U.S. Department of Energy under contract W-7405-ENG-36.


\bibliographystyle{aipproc}   

\bibliography{sample}

\IfFileExists{\jobname.bbl}{}
 {\typeout{}
  \typeout{******************************************}
  \typeout{** Please run "bibtex \jobname" to optain}
  \typeout{** the bibliography and then re-run LaTeX}
  \typeout{** twice to fix the references!}
  \typeout{******************************************}
  \typeout{}
 }

\end{document}